\documentclass[twocolumns,fleqn,usenatbib]{mnras}
\usepackage{newtxtext,newtxmath}
\usepackage[T1]{fontenc}
\DeclareRobustCommand{\VAN}[3]{#2}
\let\VANthebibliography\thebibliography
\def\thebibliography{\DeclareRobustCommand{\VAN}[3]{##3}\VANthebibliography}
\usepackage{graphicx}        
\usepackage{amsmath}        
\usepackage{dcolumn}
\usepackage{bm}
\makeatletter    
\usepackage{pifont}
\newcommand{\cmark}{\ding{51}}%
\newcommand{\xmark}{\ding{55}}
\usepackage{color,xcolor}
\newcommand{\be}{\begin{equation}}
\newcommand{\ee}{\end{equation}}
\title[Dipolar dark matter simulations]{Dipolar dark matter simulations on galaxy scales with the RAMSES code}
\author[C. Stahl et al.]{
Cl\'ement Stahl$^{1}$,
Benoit Famaey$^{1}$,
Guillaume Thomas$^{2,3}$,
Yohan Dubois$^{4}$,
Rodrigo Ibata$^{1}$,
\\
$^{1}$Universit\'e de Strasbourg, CNRS UMR 7550, Observatoire astronomique de Strasbourg, 11 rue de l'Universit\'e, 67000 Strasbourg, France\\
$^{2}$ Instituto de Astrofísica de Canarias, E-38205 La Laguna, Tenerife, Spain\\
$^{3}$ Universidad de La Laguna, Dpto. Astrofísica, E-38206 La Laguna, Tenerife, Spain  \\
$^{4}$ Institut d'Astrophysique de Paris, CNRS, Sorbonne Universit\'{e}, UMR 7095, 98bis bd Arago, 75014 Paris, France
}
\pubyear{2022}
\begin{document}
\label{firstpage}
\pagerange{\pageref{firstpage}--\pageref{lastpage}}
\maketitle
\begin{abstract}
We numerically explore on galaxy scales the Dipolar dark matter (DM) model based on the concept of gravitational polarization. This DM model has been proposed as a natural way to reproduce observed tight galactic scaling relations such as the baryonic Tully-Fisher relation and the Radial Acceleration Relation. We present a customized version of the \texttt{RAMSES} code including for the first time the dynamics of this Dipolar DM in $N$-body simulations. As a first application of this code, we check that we recover an equilibrium configuration that had been found analytically, where a low density Dipolar DM halo is at rest with respect to its central galaxy, recovering the aforementioned scaling relations. A characteristic signature of this equilibrium model is that it harbours a dynamical instability with a characteristic time depending on the Dipolar DM halo density, which we recover numerically. This represents a first step towards more involved simulations needed to test this framework, ranging from galaxy interactions to structure formation.
\end{abstract}




\section{Introduction}
The nature of the dark sector of the Universe, comprised of the putative dark matter (DM) and dark energy components, is today one of the most pressing questions in physics. Over the last decades, the cosmological picture that has emerged is that of a matter sector of the Universe dominated by cold dark matter (CDM\footnote{CDM is made of non-baryonic particles that are non-relativistic at decoupling and that, for all practical purposes in cosmology, do not interact with themselves or with baryons}) that represents $\sim 85$\% of the global matter content, itself only accounting for $\sim 30$\% of the energy budget at (present-day) redshift $z=0$. The rest of the energy budget (dark energy) is taken care of by a cosmological constant $\Lambda$, responsible for the late-time acceleration of the expansion of the Universe. 

Whilst this $\Lambda$CDM model is successful on large scales \citep{Planck:2018vyg, Gil-Marin:2020bct}, some tensions remain, both cosmologically \citep[e.g.,][]{Verde:2019ivm, Abdalla} and on the scale of individual galaxies \citep[e.g.,][]{Bullock:2017xww}. Galaxy formation and evolution happen however on scales where the complex physics of baryons can play a non-trivial role in shaping the DM distribution, through gravitational feedback: modelling precisely this complex physics and feedback is an active field of research, relying at present on subgrid models in cosmological hydrodynamical simulations \citep[e.g.,][]{Schaye, Wang, Pillepich, Hopkins, Dubois}. Currently, different simulations produce different results and the jury is still out on whether at least some of the `small-scale' problems  of $\Lambda$CDM can be addressed through feedback alone. In particular, it is highly unclear whether the apparent conspiracy between the distribution of baryons and DM in galaxies -- giving rise to tight scaling relations such as the Baryonic Tully-Fisher Relation \citep[BTFR,][]{BTFR1, BTFR2} or the Radial Acceleration Relation \citep[RAR,][]{RAR1, RAR2} and its associated diversity of rotation curve shapes \citep{diversity1, diversity2} -- can be explained naturally ~\citep[for discussions within LCDM, see, e.g.,][]{Ferrero2017,Glowacki2020,Dubois}. 

In this context, it is mandatory to explore whether the above challenges could perhaps find their root in an alteration of the fundamental nature of DM. In view of the tight correlation between the gravitational field generated by baryons and the total one in galaxies, the most direct and also most radical alternative would be that gravity has, in fact, a different behavior in galaxies and that DM is non-existent on these scales. This hypothesis, put forward almost exactly 40 years ago by \citet{Milgrom1983}, is known as Modified Newtonian Dynamics (MOND). With a simple recipe for the alteration of the gravitational law in the ultra-low acceleration regime, this hypothesis naturally reproduces most galactic observations \citep{FamaeyMcGaugh2012}, and in particular the BTFR and RAR which were actually predicted by Milgrom well before they were precisely assessed by observations. The main challenge with such an approach, however,  is to reproduce the dynamics of galaxy clusters as well as all the successes of $\Lambda$CDM on large scales, including the angular power spectrum of the CMB \citep{Planck:2018vyg} or the matter power spectrum \citep{Chabanier:2019eai}. Recent realisations of relativistic MOND theories \citep{Skordis:2020eui} address some of these problems through a Lagrangian depending on the spatial gradient squared of a scalar field as well as on its temporal derivative. The time-dependent term can then mimick the effect of DM in time-dependent configurations, and for instance reproduce the angular power spectrum of the CMB. Such bottom-up approaches however still lack a founding principle to justify the very peculiar form of the Lagrangian. Another possible approach is then to consider that  DM is basically just CDM cosmologically-speaking, but that additional fundamental medium- or long-range interactions with baryons naturally reproduce MOND in galaxies \citep[e.g.,][]{Berezhiani:2015bqa, Berezhiani2, BF1, BF2}. The observed tight galactic scaling relations such as the BTFR and the RAR would then not amount to an emergent phenomenon related to stellar feedback, but would rather have a fundamental origin, related to the very nature of DM itself. 

One of the first suggestions of this kind in the literature was that of \citet{Blanchet2007}, who proposed that galactic scale phenomenology could result from the gravitational polarization of a DM fluid carrying dipole moments aligned with the gravitational field, as we will review in detail in Sect.~\ref{sec:Theo}. This original hypothesis actually stemmed from the fact that the MOND Poisson equation proposed by \citet{Bekenstein:1984tv} is in manifest analogy with the first of Maxwell's equations in matter, describing the electric field inside a dielectric medium, with the dielectric coefficient being in analogy with the so-called `interpolating function' of MOND. In a non-linear medium, the dielectric coefficient can be a function of the magnitude of the electric field, and it always screens the electric charges, thereby reducing the electric field within the dielectric. However, contrary to electrostatics, `gravitational charges' of the same sign (i.e., simply, masses) always attract each other: therefore, in a `digravitational' medium the digravitational coefficient should produce an anti-screening of the ordinary masses by the polarization masses, thereby enhancing the gravitational field within such a medium.

Subsequently, \citet{Blanchet:2008fj} proposed a covariant action for describing phenomenologically the dynamics of this DM medium endowed with a dipole moment vector, and polarizable in a gravitational field. This model was then refined and its phenomenology was analytically studied in detail in \citet{Blanchet_2009}, notably showing that it does recover $\Lambda$CDM at linear
perturbation order. Another variant of the model was later proposed in \citet{Bernard} to actually allow for an easier microphysics interpretation of the dipole moment, and it was then refined to avoid ghost instabilities at linear order in \citet{BlanchetHeisenberg2015}, and shown to be an acceptable effective field theory below the strong coupling scale \citep{BlanchetHeisenberg2017}. 

Numerical simulations of galaxies embedded in such a dipolar DM medium have however never been carried out. Here, we propose to implement for the first time the equations of motion of dipolar DM in the \texttt{RAMSES} code \citep{Teyssier:2001cp} to perform the first-ever galaxy simulations in this context. Our ambition is limited in scope in this first exploratory paper: we mainly set ourselves the task of checking that the analytical results previously obtained are recovered in numerical simulations. This will serve as a first step toward the development of  full simulations of galaxy interactions, and later structure formation, in this context. 

For the present work, we start from the original covariant formulation of \citet{Blanchet:2008fj} and \citet{Blanchet_2009}: despite its lack of microphysical interpretation in terms of actual DM particles, this model adequately describes the effective dynamics of a DM fluid endowed with a dipole moment vector. The model is reviewed in Sect.~\ref{sec:Theo}, and we then present our numerical setup in Sect.~\ref{sec:num}. Results are presented in Sect.~\ref{sec:res} and conclusions and perspectives are discussed in Sect.~\ref{sec:ccl}.

\section{Dipolar dark matter: theoretical framework}
\label{sec:Theo}
The dipolar DM hypothesis is a proposition that aims to explain the observed phenomenology at galactic scales, that is usually well described by MOND, as the consequence of a new fundamental property of the DM medium. Note that this dipolar DM framework will {\it not} necessarily always reproduce the MOND phenomenology, and could therefore in principle also account for deviations from it. The hypothesis is motivated by the form of the MOND Poisson equation \citep{Bekenstein:1984tv},
\be
\vec{\nabla} \cdot (\mu \vec{g}) = - 4 \pi G \rho_{\rm b} ,
\ee
where $\rho_{\rm b}$ is the baryon density, $\vec{g}$ is the gravitational acceleration, and $\mu(g)$ is the so-called `interpolating function' which approaches $g/a_0$ when $g \ll a_0$ and 1 when $g \gg a_0$, where $g = |\vec{g}|$ is the norm of the gravitational acceleration and $a_0 \simeq 10^{-10} \, {\rm m} \, {\rm s}^{-2}$ is an acceleration constant. As noted by \citet{Blanchet2007}, this equation is in manifest analogy with the first of Maxwell's equations in matter, describing the electric field inside a dielectric medium,
\be
\vec{\nabla} \cdot (\mu_e \vec{E}) = 4 \pi \rho_e ,
\ee
where $\rho_e$ is the free charge density, $\vec{E}$ is the electric field, and $\mu_e = 1 + \chi_e$ is the dielectric coefficient, where $\chi_e(E)$ is the electric susceptibility that can be a function of the norm of the electric field in a non-linear medium. By analogy, one can therefore rewrite the MOND interpolating function as a `digravitational' coefficient $\mu(g) = 1 + \chi(g)$, and rewrite the generalized Poisson equation of MOND as
\be
\nabla^2 \Phi = 4 \pi G (\rho_{\rm b} -  \vec{\nabla} \cdot \vec{\Pi}),
\label{eq:MOND}
\ee
where $\Phi$ is the gravitational potential such that $-\vec{\nabla} \Phi = \vec{g}$, and $\vec{\Pi} = - (\chi \vec{g})/(4 \pi G)$. For instance, for the so-called `simple' interpolating function of MOND \citep{FamaeyBinney, Gentile2011}, the corresponding form of the $\chi$-function is $\chi(g) = -a_0 / (a_0 + g)$.

The term $-  \vec{\nabla} \cdot \vec{\Pi}$ in Eq.~\ref{eq:MOND} is usually called the `phantom DM' density in the MOND context: it is in this context a mathematical erzatz representing the DM density that would produce the MOND gravitational potential in the context of Newtonian gravity \citep[e.g.,][]{Milgrom1986,Oria}. In that context, it is a mathematical artefact, a fictitious mass term, but the idea of the dipolar DM model is to make this term a {\it true} source of the gravitational field, without modifying the Newtonian Poisson equation, by considering it to be the mass density $\rho_{\rm pol}$ of the polarization masses associated with the polarization field $\vec{\Pi}$ in a `digravitational' DM medium. In order to reproduce the MOND phenomenology, one simply needs the polarization field $\vec{\Pi}$ to be aligned with the gravitational field $\vec{g}$, with the right gravitational susceptibility $\chi$.

\subsection{The action and equations of motion}

To effectively realize such a DM model, \citet{Blanchet:2008fj} and \citet{Blanchet_2009} proposed a covariant action for a polarizable DM fluid. Let $x^{\mu}$ be the coordinates of spacetime, $\tau$ the proper time associated with the metric $g_{\mu \nu}$, and $u^{\mu} \equiv \frac{d x^{\mu}}{d \tau}$ the usual proper 4-velocity of the fluid. The dipolar DM fluid is described by a mass current $J^{\mu} \equiv \sigma u^{\mu}$ where $\sigma$ is the rest mass density. As the velocity is time-like, one has $\sigma^2 = - J_{\mu} J^{\mu}$, and as expected the 4-current is conserved $\nabla_{\mu} J^{\mu}=0$, where $\nabla_{\mu}$ is the canonical covariant derivative associated with $g_{\mu \nu}$. As the fluid is polarizable, it also carries a dipole moment vector $\xi^{\mu}$. It has dimension of length, and its corresponding polarization 4-vector reads: $\Pi^{\mu} \equiv \sigma \xi^{\mu}$. As in standard electromagnetism, the polarization vector is a coarse grained quantity which impacts the master equations of motion. The (complicated) small scale behavior of the individual microscopic components of the fluid are \textit{not} modelled. In this sense, the theory that we consider here is an {\it effective} theory. In the following, $\Pi^{\mu}$ or equivalently $\xi^{\mu}$ will be considered as an independent variable from $J^{\mu}$, and is an \textit{internal} degree of freedom of the DM fluid. This is to be contrasted with the typical vector-tensor theories where the extra vector is coupled to the metric (e.g., \citealt{Zlosnik}, see \citealt{Heisenberg:2018vsk} for a review and \citealt{Gomez:2022okq} for a recent take on it).

Following \citet{Blanchet_2009}, the action describing our polarized fluid reads in natural units:

\be 
\label{eq:action}
S = \int d^4 x \sqrt{-g} \left[- \sigma + J_{\mu}\dot{\xi^{\mu}}- \mathcal{W} \right],
\ee
where $\mathcal{W}$ is a potential depending on the norm of the space-like projection of the polarization vector $\Pi^{\mu}$ perpendicular to the 4-velocity of the fluid. A suitable choice of the potential will allow one to recover the MOND phenomenology on galactic scales. Our propagating degrees of freedom are the matter fluid and its polarization vector. The dot denotes here the Lagrangian derivative along the fluid trajectory, that is $\dot{\xi^{\mu}}=u^{\nu}\nabla_{\nu} \xi^{\mu}$. As $J^{\mu}$ is time-like, the presence of the coupling between the polarization and the matter current restricts the polarization to be space-like since a time-like part would be a pure divergence term and would not propagate. This is to be contrasted with Einstein-aether \citep{Jacobson:2007veq} and TeVes-like theories \citep{Bekenstein:2004ne} that consider a new time-like vector field. Throughout the rest of this article, we will only consider the space-like projections of our new internal variables perpendicularly to the 4-velocity  ($\vec{\xi}$ and $\vec{\Pi}$ whose norm will be denoted $\xi$ or $\Pi$).

Aiming at implementing the dipolar DM framework into a N-body experiment, we now consider the non-relativistic limit of Eq.~\eqref{eq:action} which after explicitly expanding the coupling $J^{\mu}\dot{\xi^{\mu}}$ gives for the matter action of DM \citep{Blanchet_2009}:
\be 
\label{eq:action}
S = \int d^4 x \left[\rho_{\rm dm} \left(\frac{\vec{v}^2}{2}-\Phi+\vec{g} \cdot \vec{\xi} + \vec{v} \cdot \frac{d \vec{\xi}}{dt} \right)- \mathcal{W}(\Pi) \right],
\ee
where $\vec{v}$ is the fluid velocity, $\Phi$ is the usual Newtonian gravitational potential, $\vec{g}=-\vec{\nabla} \Phi$ the usual Newtonian gravitational force, and $d/dt$ is the Lagrangian derivative. Note also that, from here, our notation differs from the one of \cite{Blanchet_2009} as we denote the monopolar part of the DM density as $\rho_{\rm dm}$ (called $\sigma^*$ in \citealt{Blanchet_2009}). 

Varying the action with respect to its degrees of freedom, one gets the equations of motion
\be
\label{eq:mod force}
\frac{\text{d} \vec{v}}{dt}= -\vec{\nabla} \Phi-\vec{F}_{\text{int}},
\ee
where
\be
\label{eq:Fint}
\vec{F}_{\text{int}}=\frac{\vec{\Pi}}{\Pi} \mathcal{W}', 
\ee
and
\be
\label{eq:newfield}
\frac{\text{d}^2 \vec{\Pi}}{\text{d}t^2}=\rho_{\rm dm}\vec{F}_{\text{int}} +  \vec{\nabla}\left(\mathcal{W}-\Pi\mathcal{W}' \right) + (\vec{\Pi} \cdot {\vec{\nabla}}) \vec{\nabla} \Phi .
\ee

Note that, unlike standard DM, the polarization fluid is corrected by a new internal force whose intensity depends on the first derivative of the potential, $\mathcal{W}'$. This shows that in a General Relativistic setup the motion of DM is not geodesic. 

Finally, the gravitational field equation is obtained by varying the action with respect to $\Phi$, together with the usual gravitational part of the action in the weak-field limit $S_{\rm grav} = \int d^4 x (\nabla \Phi)^2/(8 \pi G)$, and with the baryon contributions to the matter action, giving for the Poisson equation:
\be
\label{eq:mod grav}
\nabla^2 \Phi=4 \pi G \left(\rho_{\rm b}+\rho_{\rm dm} - \vec{\nabla} \cdot{} \vec{\Pi}\right).
\ee
Remarkably, if the dipolar DM fluid clusters weakly on galaxy scales in the non-linear regime, i.e. with $\rho_{\rm dm} \ll \rho_{\rm b}$ such that $\rho \approx \rho_{\rm b}$, this is precisely the form of the MOND Poisson equation (Eq.~\ref{eq:MOND}), as long as $\vec{\Pi} = - \frac{\chi(g) \vec{g}}{4 \pi G}$.

\subsection{The potential}
\label{sec:pot}
The next crucial aspect is to understand how the polarization of the DM medium can allow the polarization field to have the desired dependence on the gravitational field $g$. \citet{Blanchet:2008fj} and \citet{Blanchet_2009} showed that, at linear perturbation order in cosmology, the dipolar DM model is essentially identical to what is expected in the $\Lambda$CDM context. However, it was also argued that the dynamics should be vastly different in the non-linear regime because of the internal force counterbalancing the effect of gravity. It was indeed shown that an exact solution of the equations governing the dynamics of dipolar DM exists when the dipole moments are in equilibrium at rest, and the DM fluid has zero net velocity in its halo reference frame. In this equilibrium configuration, the internal force therefore precisely counterbalances gravity, and the DM particles that are not at rest would simply fly away from the galactic halo. It was thereby heuristically inferred that the DM density contrast in a typical galaxy at low redshift should be very small, with a typical density not much above the critical density, whilst the internal force should be basically equal to the gravitational one, in direction and magnitude, i.e.
\be
\label{eq:Pitog}
\mathcal{W}'(\Pi) = g.
\ee
In the deep-MOND regime ($g \ll a_0$), we know that we need $\vec{\Pi} = - \frac{\chi(g) \vec{g}}{4 \pi G}$ with $\chi(g) = (g/a_0)-1$, hence
\be
\label{eq:expand}
\vec{\Pi} = \frac{\vec{g}}{4 \pi G} \left( 1 - \frac{g}{a_0}\right).
\ee
Taking into account that the two vectors point in the same direction and solving this equation for $g$, then Taylor-expanding the solution around zero, leads to the requirement that
\be
\label{eq:Taylor_g}
g = \mathcal{W}'(\Pi) = 4 \pi G \Pi + \frac{(4 \pi G \Pi)^2}{a_0} + \mathcal{O}(\Pi^3).
\ee

Therefore, the explicit expression for the potential in the deep-MOND limit proposed in \citet{Blanchet_2009} was:
\be
\label{eq:polpot}
\mathcal{W}= \frac{a_0^2}{8\pi G} + 2\pi G \Pi^2+ \frac{(4 \pi G)^2}{3 a_0} \Pi^3 + \mathcal{O}(\Pi^4).
\ee
This potential, valid for the deep-MOND regime, can in principle be adapted to any form of the interpolating function. Note that the first term is chosen to have the same order of magnitude ($\sim a_0^2$ when considering the expansion in powers of $\Pi/a_0$) and units as the other ones, and can naturally play the role of the cosmological constant $\Lambda \sim a_0^2$ as the zero point of the dipolar DM potential in this context.

The smaller the value of $\Pi/a_0$, i.e. the deeper one is in the MOND regime, the better the Taylor-expanded version of the potential is accounting for the MOND dynamics. However, the absence of higher order terms can lead to small deviations which could lead to some unexpected effects in our numerical experiments. Therefore, for later use, we also lay down here the exact potential obtained by solving Eq.~\eqref{eq:expand} and integrating without Taylor expanding, but keeping the same 0-th order term as above:
\be
\label{eq:exactW}
\mathcal{W}_{e}= \frac{a_0^2}{8\pi G}\left[\frac{4 \pi G \Pi}{a_0} +\frac{1}{6}\left(1-\frac{16 \pi G \Pi}{a_0} \right)^{3/2}+\frac{5}{6} \right].
\ee
Anticipating our numerical results, this exact deep-MOND version of the potential will be useful to control numerical errors.

\section{Numerical setup}
\label{sec:num}

A static spherically symmetric DM fluid with $\vec{v}=\vec{0}$ (namely, no tangential nor radial motions) is an equilibrium configuration, as previously shown analytically by \citet{Blanchet_2009}. However, it is an unstable equilibrium, but the characteristic time of the instability was shown to be of the order of a Hubble time. The {\it main goal} of the present paper will be to implement the equations of dipolar DM in the \texttt{RAMSES} code \citep[in the spirit of][in the MOND context]{Lughausen}, and use the analytical solution found by \citet{Blanchet_2009} as a very first application of the code. This equilibrium solution also relies on the {\it weak clustering hypothesis}, namely that once the medium polarizes itself in the non-linear regime of structure formation, the internal force counteracts gravity, and allows most particles not at rest to escape. Since the internal force of the dipolar DM fluid almost exactly compensates for gravity, an initial population of DM particles with a velocity dispersion of $\sim 30$~km/s$~\simeq 30$~kpc/Gyr would leave within 1~Gyr only a very low monopolar density around the galaxy, almost at rest with respect to its own frame. This reasonable hypothesis in the dipolar DM context will need to be backed by further simulations of structure formation and is beyond the scope of the present work. Thus in all simulations presented in this work, the dark matter has zero initial velocity: $\vec{v}=\vec{0}$. Moreover, $\dot{\vec{\Pi}}(t=0)=0$ as per the equilibrium solution of \citet{Blanchet_2009}.

\subsection{Initial Conditions for the polarization}

In the previous section, we have shown how to devise a potential $\mathcal{W}$ with the desired behaviour in the deep-MOND regime. 
Defining $\vec{g}_n \equiv - \nabla \Phi_N$, the force a test particles would feel in a Newtonian setup with baryons only, the deep-MOND regime implies $g = \sqrt{g_n a_0}$. As an initial condition for $\vec{\Pi}$ in this regime, we impose Eq.~\eqref{eq:expand} together with $\dot{\vec{\Pi}}(t=0)=0$. Such a choice ensures that the dipolar DM is in equilibrium at rest. We also ensured that the monopolar DM density is low, hence that the weak clustering hypothesis applies, and we will check later in our numerical experiment that we are indeed in the deep-MOND regime. Our goal is to simulate the equilibrium solution investigated analytically in previous work: for this, we will be setting up a spherical galaxy, for both baryons and dipolar DM.

\subsection{Time evolution}
With the initial conditions for the polarization vector chosen in the previous subsection, the initial potential is identical to the deep-MOND regime. However, one specificity of the dipolar DM is that the polarization vector has a specific dynamics given by Eq.~\eqref{eq:newfield}. For the experiment we are planning to conduct in the context of the static equilibrium solution, and in order to follow the exact same assumptions as in the analytical case, we follow the approximation made in Appendix A of \cite{Blanchet:2008fj}, neglecting all but the first term of the right hand side of Eq.~\eqref{eq:newfield}:
\be
\label{eq:dynPi}
\frac{\text{d}^2 \vec{\Pi}}{\text{d}t^2}=\rho_{\rm dm}\vec{F}_{\text{int}}.
\ee
 For our purpose of spherical symmetry, it is a good approximation and allows to check the setup and results of \citet{Blanchet:2008fj} and \citet{Blanchet_2009}. In follow-up works, we will need to take into account the non-spherical terms.  

Following the usual update on the position and velocity of particles in \texttt{RAMSES} \citep{Teyssier:2001cp} which is done by a leapfrog scheme, we implemented the time-evolution of the polarization vector field with a second leapfrog that allows for symplectic dynamics also for the polarization vector $\vec{\Pi}$. For a given timestep $\Delta t$ that we choose equal to the particles time step update\footnote{Here, we also assume the weak clustering hypothesis: the time step $\Delta t$ is chosen as the minimum of the Courant conditions and freefall times for the polarization $\vec{\Pi}$ and the baryons. However the two timescales related to $\vec{\Pi}$ (the Courant like condition $\propto \Pi/\dot{\Pi}$ and the freefall time $\propto 1/\sqrt{\rho_\text{dm}}$) are much larger than the freefall baryon time scale $1/\sqrt{\rho_\text{b}}$, provided that we are in the weak clustering hypothesis.}
\begin{eqnarray}
& \dot{\vec{\Pi}}(t+\Delta t/2)=\dot{\vec{\Pi}}(t) + \vec{F}_{\text{int}}(t) \rho_{\rm dm} \frac{\Delta t}{2} \\
& \vec{\Pi}(t+\Delta t)=\vec{\Pi}(t) + \dot{\vec{\Pi}}(t+\Delta t/2) \Delta t \\
& \dot{\vec{\Pi}}(t+\Delta t)=\dot{\vec{\Pi}}(t+\frac{\Delta t}{2} ) + \vec{F}_{\text{int}}(t+\Delta t) \rho_{\rm dm} \frac{\Delta t}{2}
\end{eqnarray}
We store the value of $\vec{\Pi}(t)$ (as well as those of $\dot{\vec{\Pi}}$, $\rho_{\rm dm}$ and $F_{\rm int}$) on the cells of the \texttt{RAMSES} grid to be able to evolve the gravitational sector. Assuming the cells are indexed $i$, $j$, $k$, all those quantities are centered in $i$, $j$, $k$ while the divergence in equation \eqref{eq:mod grav} is calculated using the standard finite difference on neighbouring cells: 
\be
\begin{split}
& \vec{\nabla} \cdot{} \vec{\Pi}=\left[(\vec{\Pi}_{i+1,j,k}-\vec{\Pi}_{i-1,j,k}) + \right. \\ & \left. (\vec{\Pi}_{i,j+1,k}-\vec{\Pi}_{i,j-1,k}) + (\vec{\Pi}_{i,j,k+1}-\vec{\Pi}_{i,j,k-1})\right]/(2\text{d}x),
\end{split}
\ee
with $\text{d}x$ is the spacing of the fine \texttt{RAMSES} grid.

As discussed in \citet{Blanchet_2009}, the time evolution leads to an instability that destabilizes the galaxy on a time scale of the order of the free-fall time $\tau \propto 1/\sqrt{\rho_\text{dm}}$, which is of the order of a Hubble time for densities around the critical density. Our goal will now be to test this analytical prediction with our numerical scheme.

\subsection{Simulating a dwarf galaxy with a King profile}

\begin{table*}
\begin{center}
\caption{Summary of the different simulations carried out in this paper.
  The columns give the simulation identifier, the dark matter density at the distribution radius in units of the critical density (Using $H_0=70$ km/s/Mpc, the critical density reads $\rho_c=136 \text{ M}_\odot/\text{kpc}^3$), the radius of the dark matter distribution in kpc, the number of particles of dark matter, whether dark matter is allowed to move, the baryon density in units of the critical density, the total mass of the baryons in solar masses, the number of baryon particles, the tidal radius of the King model in kpc, the number of particles, their mass in solar masses, whether baryons are allowed to move and whether the polarization dynamics (given by equation \eqref{eq:newfield}) is turned on. The size of the simulated box is always $200$ kpc, with a grid of $128^3$. We allowed for 12 refinement levels thus leading to an effective resolution of 49 pc.}
\begin{tabular}{ l || c | c | c |  c | c | c || c | c | c | c | c | c }
\hline\hline  
Simulation & $\rho_{\text{dm}}$ & $r_{\text{dm}}$ & $N_{\text{pcl}}$ & $m_{\text{pcl}}$ & DM & $\vec{\Pi}$  & $\rho_{\text{b}}$ & $M_{\text{b}}$ &  $r_t$ & $N_{\text{pcl}}$ & $m_{\text{pcl}}$  & baryon    \\
 &  & (kpc)  &  & (M$_\odot$) & moves? &  moves?  &  & (M$_\odot$) &  (kpc) & & (M$_\odot$)  & moves?   \\
\hline 

$\text{K}_{\text{dm}}$ & 10 $\rho_c$  & 20 & $4 \times 10^6$ &11 & \cmark  & \xmark & $10^5$  $\rho_c$ & $1.1 \times 10^8$ & 2.7 & $10^5$ & $1.1 \times 10^3$ & \xmark    \\

$\text{K}_{\text{b}}$ & 10 $\rho_c$  & 20 & $4 \times 10^6$ &11 & \cmark   & \xmark & $10^5$  $\rho_c$ & $1.1 \times 10^8$ & 2.7 & $10^5$ & $1.1 \times 10^3$ & \cmark   \\

$\text{K}_{10}$ & 10 $\rho_c$  & 20 & $4 \times 10^6$ &11 & \cmark & \cmark  & $10^5$ $\rho_c$ & $1.1 \times 10^8$ & 2.7 & $10^5$ & $1.1 \times 10^3$ & \cmark    \\

$\text{K}_{1}$ & $\rho_c$  & 20 & $4 \times 10^6$ &1.1 & \cmark & \cmark  & $10^5$ $\rho_c$ & $1.1 \times 10^8$ & 2.7 & $10^5$ & $1.1 \times 10^3$ & \cmark    \\

$\text{K}_{4}$ & 4 $\rho_c$  & 20 & $4 \times 10^6$ &4.6 & \cmark & \cmark  & $10^5$ $\rho_c$ & $1.1 \times 10^8$ & 2.7 & $10^5$ & $1.1 \times 10^3$ & \cmark    \\

$\text{K}_{7}$ & 7 $\rho_c$  & 20 & $4 \times 10^6$ &8.0 & \cmark  & \cmark &  $10^5$ $\rho_c$ & $1.1 \times 10^8$ & 2.7 & $10^5$ & $1.1 \times 10^3$ & \cmark    \\

\hline
\end{tabular}
\label{table:simtable}
\vspace{-5mm}
\end{center}
\end{table*}

Table \ref{table:simtable} sums up the parameters used in all our simulations. In particular, the right-hand part of the table describes the distribution of baryons, which represent a spherical low-surface brightness dwarf galaxy. 

\begin{figure*}
        \centering
        \includegraphics[width=\textwidth]{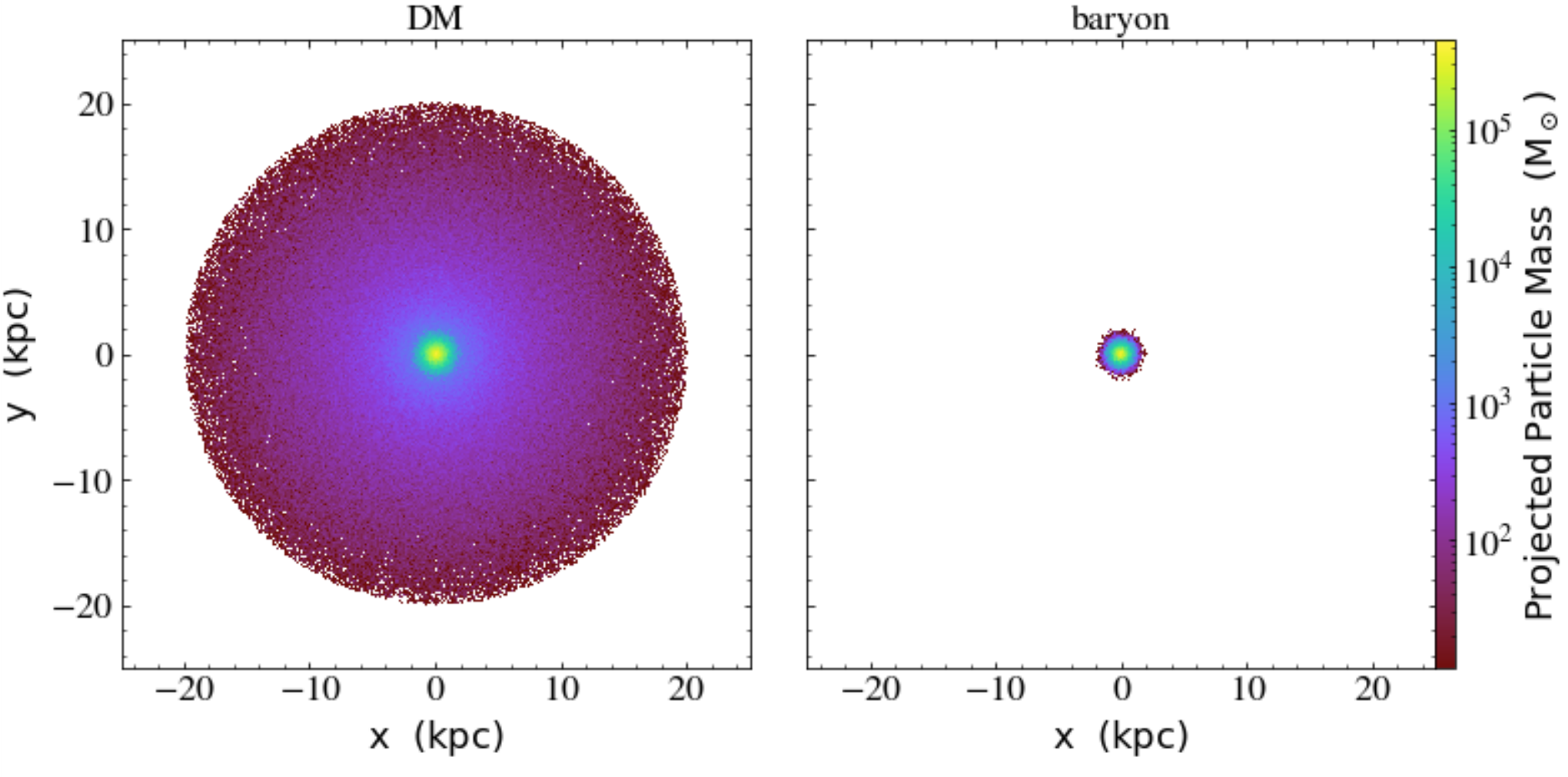}
        \caption{\label{fig:CI} A particle projection of our initial configurations for the dark matter and the baryons in the simulation $K_{10}$. The total box length is 200 kpc. In the left panel, we display the uniform sphere of density $10 \rho_c$. Due to projection effects, more dark matter particles are projected at the center of the box but the sphere is still uniform. In the right panel, the baryonic core consists of a King profile, corresponding to a total mass of $1.1 \times 10^8 \text{ M}_{\odot}$}
\end{figure*}

The dark matter is represented by a uniform sphere of $4 \times 10^6$ particles of polarized dark matter with a density\footnote{Densities are in units of the critical density of the universe: $\rho_c=136 \text{ M}_\odot/\text{kpc}^3$.} which can be as low as the critical density and with zero initial velocity: $\vec{v}=\vec{0}$. The initial state of the polarization field is here given by Eq.~(\ref{eq:expand}) with $g = \sqrt{g_n a_0}$. We present this configuration in the left panel of figure \ref{fig:CI}. We first tested that the limiting radius of the DM sphere did not affect the results in the case of a central point mass representing the baryons. We tried two different limiting radii $r_{\rm dm}=70$~kpc and $r_{\rm dm}=20$~kpc, getting an identical behaviour for the dynamics of the DM and its polarization field. Therefore, we chose $r_{\rm dm}=20$~kpc for all the following simulations.

\begin{figure}
        \centering
        \includegraphics[width=0.47\textwidth]{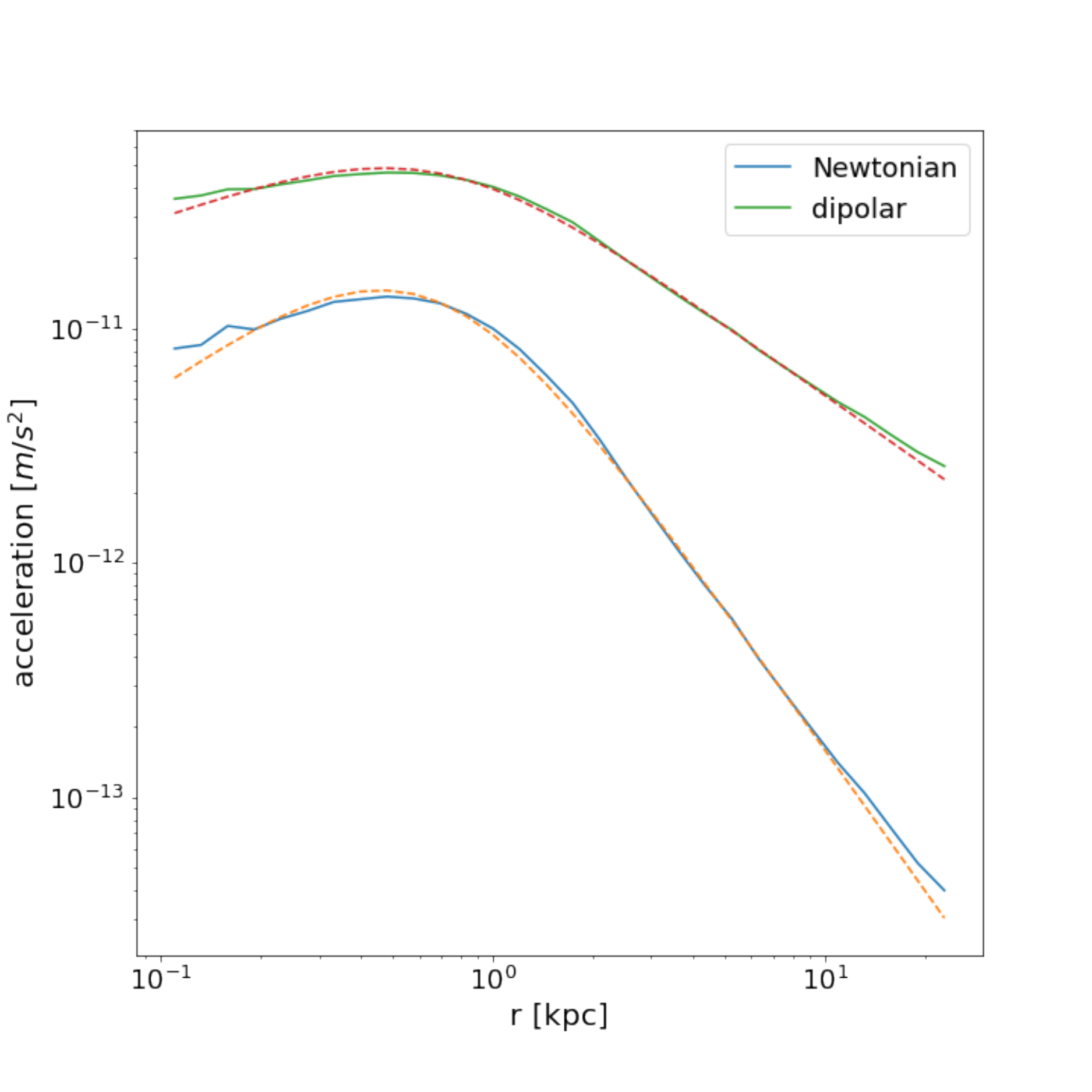}
        \caption{\label{fig:accel} The initial acceleration profile for the simulation $\text{K}_{10}$. The total box length is 200 kpc. The baryonic core consists of a King profile, corresponding to a total mass of $1.1 \times 10^8 \text{ M}_\odot$. The new force introduced by the dipolar DM indeed scales as $r^{-1}$ as expected from the MONDian dynamics in its asymptotic regime. We represent the theoretical prediction in dashed lines and the measurements from our N-body experiments in plain lines. Note that at large and small $r$, the measurements starts to differ from the prediction, due to numerical discretization effects of the grid and the number of particles used to model the constant density sphere of DM. }
\end{figure}

The (collisionless) baryons of the dwarf galaxy are represented by a self-consistent King model in the deep-MOND regime of baryonic mass $1.1 \times 10^8\, \rm M_\odot$. The King model, as described in Section 4.3 of \citet{BT08}, is a lowered isothermal model described by a distribution function depending on the binding energy, a characteristic density and a central velocity dispersion parameter $\sigma$. It hence has a density profile which is proportional to the relative binding potential $\psi$:
\be
\rho \propto e^{\psi/\sigma^2} \text{erf} \left(\frac{\sqrt{\psi}}{\sigma} \right) -  \sqrt{\frac{4 \psi}{\pi \sigma^2}} \left(1+ \frac{2 \psi}{3 \sigma^2} \right),
\ee
where erf represents the error function. The Poisson equation can then be solved inside-out, and the chosen central value of $\psi$ parametrizes the limiting radius $r_t$ at which the density falls to zero. The concentration of a King model is given by $C=\log(\sqrt{4 \pi G \rho_0}r_t/3 \sigma)$ where $\rho_0$ is its central density. In our case, the MOND Poisson equation is solved instead of the Newtonian one: the explicit construction and the validation of such a MONDian King model are given in \citet{Thomas2017}\footnote{The program used to generate the baryonic initial conditions is available at \url{https://github.com/GFThomas/MOND}.}. For this work, we will use the parameters $\sigma= 25$~km~s$^{-1}$ for the central velocity dispersion, a limiting radius $r_t= 2.38$~kpc, and a concentration $C=0.376$, leading to a half-light radius $r_h= 0.7764$~kpc. Our model will be numerically represented by $10^5$ particles of $1146\,\rm M_\odot$. It leads to a baryonic configuration that is in the deep-MOND regime everywhere, as one can inspect from  figure \ref{fig:accel}. In the right panel of figure \ref{fig:CI}, we present a graphical representation of this initial configuration.

We ran three types of simulations with this King profile: one in which the baryons are frozen, letting only the DM and its polarization field evolve with time, a second one in which the baryons are live but the polarization field is frozen, and a third one in which the DM, its polarization field {\it and} the baryons are live. For all simulations, we chose a box length of 200 kpc discretized on a $128^3$ coarse grid. We allowed for 12 levels of refinement thus leading to an effective resolution of 49 pc. In all cases, we considered only the particle sector of \texttt{RAMSES}: in other words, we assume that the baryonic matter is in the form of (collisionless) stars. We modified the gravity sector of \texttt{RAMSES} so that baryons and DM feel at all times the gravity of baryons, of the monopolar DM density {\it and} of the dipolar $\nabla \cdot \vec{\Pi}$ term. The Poisson equation is solved using the conjugate gradient method. We imposed reflexive boundary conditions for $\vec{\Pi}$ and Dirichlet ($\Phi=0$ on the edges of the simulation box) condition for the potential.

\section{Results}
\label{sec:res}
Our goal in this paper is two-fold: first construct, under the simple hypothesis described in the previous section, equilibrium configurations for galaxies and their halo; second, explore numerically the theoretically predicted instability that is a specific signature of the dipolar DM model. 

To do so, we first performed several checks that the configuration that we initially impose is stable when turning off the dynamics of the polarization, and then we characterized the instability that arises when the dynamics of the polarization field is turned on. We did this with both the baryons being frozen and with the baryons being live.

\subsection{Stability of the dark matter halo: frozen baryons}
\label{sec:stabDM}
As a first step, we do not allow the baryons to move and turn off the dynamics of $\Pi$, given in equation \eqref{eq:newfield}. This simulation corresponds to $\text{K}_{\text{dm}}$ in Table \ref{table:simtable}. We find that the dipolar DM halo is stable on galactic ($\sim 100$ Myr) and cosmological ($\sim 10$ Gyr) times.

In figure \ref{fig:accel}, we present the Newtonian and dipolar contributions to the gravitational acceleration, that scale as $r^{-2}$ and $r^{-1}$.  This shows that we are indeed in a configuration where a MOND-like force would hold together the baryons if they were allowed to move. The DM particles, on the other hand, feel a zero net force and are thus at rest.

One interesting aspect about the potential $\mathcal{W}(\Pi)$  of Eq.~(\ref{eq:polpot}) is that, following \citet{Blanchet_2009}, it is expressed in powers of $\Pi/a_0$. Therefore, there will be a small difference between the gravitational force generated by $-\nabla \cdot \vec{\Pi}$ (with $\Pi$ determined by Eq.~(\ref{eq:expand}) in our initial conditions) and the internal force $\vec{F}_{\text{int}}$ computed from the potential of Eq.~\eqref{eq:polpot} expressed as a Taylor series in $\Pi/a_0$ with a cut-off at order 3. The smaller the value of $\Pi/a_0$, i.e. the deeper one is in the MOND regime, the better that approximation. Starting from Eq.~\eqref{eq:Fint}, $\vec{g}-\vec{F}_{\text{int}}$ can be corrected by higher order terms of the potential proportional to $ \vec{g_n} \sqrt{g_n/a_0}$. These higher order corrections to the equilibrium solution $\vec{v}=\vec{0}$ are totally negligible in the deep-MOND regime but interestingly, we could still measure them in this idealized case. We have found a perfect match with the analytical prescription from an analytical expansion to higher order terms of Eq.~(\ref{eq:Taylor_g}), thereby validating the computation of the forces in our code. 

In order not to confuse those higher order effects with the other physical effects we want to highlight in this paper, we only consider in this paper the exact potential $\mathcal{W}_e$ given in Eq.~\eqref{eq:exactW}.

\subsection{Stability of the baryonic profile: frozen polarization}
\label{sec:instab}
Now, following the configuration $\text{K}_{\text{b}}$ of Table \ref{table:simtable}, we let evolve the baryonic King profile. But we turn off the dynamics of $\vec{\Pi}$, given in Eq.~\eqref{eq:newfield}, assuming we remain in a static spherical configuration. On galactic time scales, we report that the baryonic King profile is very stable. We ran our numerical experiment for several Hubble times ($30$ Gyr) and did not find any instability. This was expected: the baryons remain stable in the MONDian gravitational potential that we have constructed initially. Checking this gave us confidence that our initial numerical setup was not swamped by numerical or dynamical instabilities that would be degenerated with our next non-trivial check involving physics proper to the dipolar DM model. 

\subsection{Everything live: instability of the polarization}
A unique feature of the dipolar DM model is the dynamics of the polarization field that we now study in detail, following the configuration $\text{K}_{\text{10}}$ of Table \ref{table:simtable}. 

\begin{figure*}
        \centering
        \includegraphics[width=\textwidth]{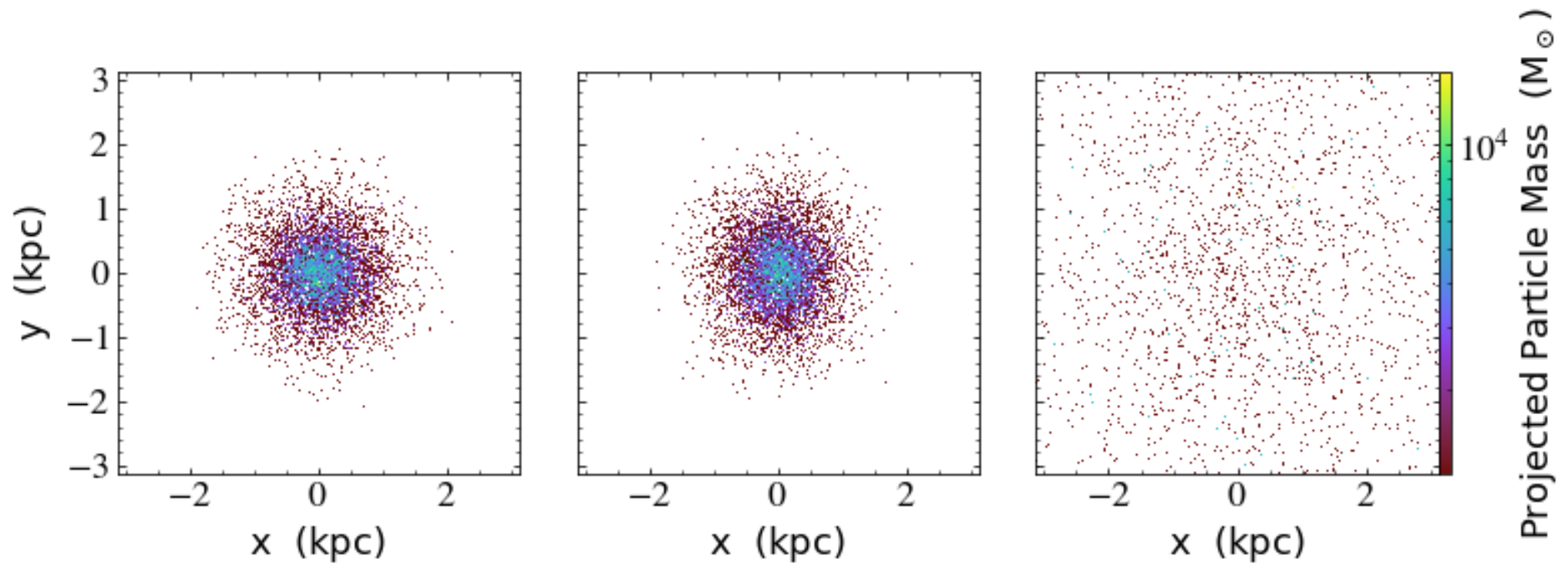}
        \caption{\label{fig:instaProjec} Particle projection of the baryons present in the simulation $K_{10}$ at three different times: 0 Gyr, 1.5 Gyr and 13 Gyr of our King model. At galactic time scales (left and middle panel), the King model remains stable, even though it has its own dynamics for each baryonic particle. At cosmological time scales, in the right panel, the dynamical instability triggered by the time evolution of the polarization vector dominates the motion of the baryons and the baryonic core has totally dissolved.}
\end{figure*}

Figure \ref{fig:instaProjec} shows snapshots of the evolution of the baryons. The first two snapshots show no significant evolution, demonstrating that on a few dynamical time scales (1.5 Gyr), our galaxy remains stable. However letting our galaxy evolve some more Hubble times (13 Gyrs), it dissolves: most of the baryons escaped from the center of the galaxy.

\begin{figure}
        \includegraphics[width=0.47\textwidth]{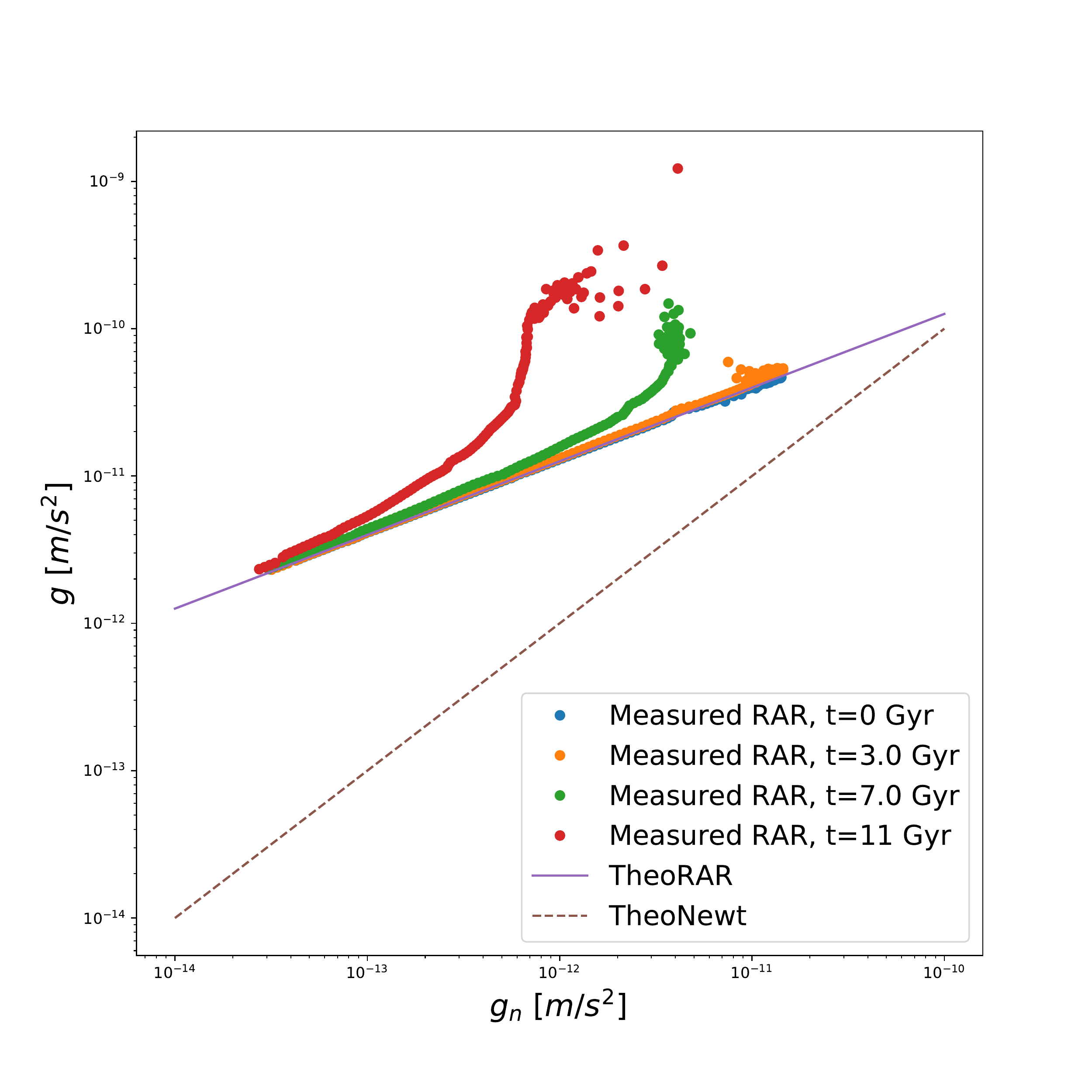}
        \caption{\label{fig:RAR} The time evolution of the Radial Acceleration Relation (RAR) for the simulation $K_{10}$. It shows that, the dipolar DM model recovers the theoretical (MOND) line with a great agreement at galactic times scales. The dashed green line represent the linear (Newtonian) relation. A departure from MOND occurs once the instabilities on $\Pi$ have developed, the RAR gets deformed as the baryons were ejected from the central (stronger field) region. Note that the RAR remains stable for longer in simulation $K_1$.}
\end{figure}

We present the RAR in figure \ref{fig:RAR} for the simulation $K_{10}$. At initial times, we recover exactly the deep-MOND regime: the RAR is then simply given by $g_{\text{tot}}=\sqrt{g_N \times a_0}$. As the dynamical instability develops, the RAR starts to deform as $\Pi$ grows exponentially, and impacts the whole gravitational sector. The Newtonian gravity decreases in the stronger field regime as the center of the galaxy gets emptied, but the the total acceleration now exceeds the deep-MOND one.

\begin{figure*}
        \centering
        \begin{tabular}[t]{cc}
        \includegraphics[width=0.47\textwidth]{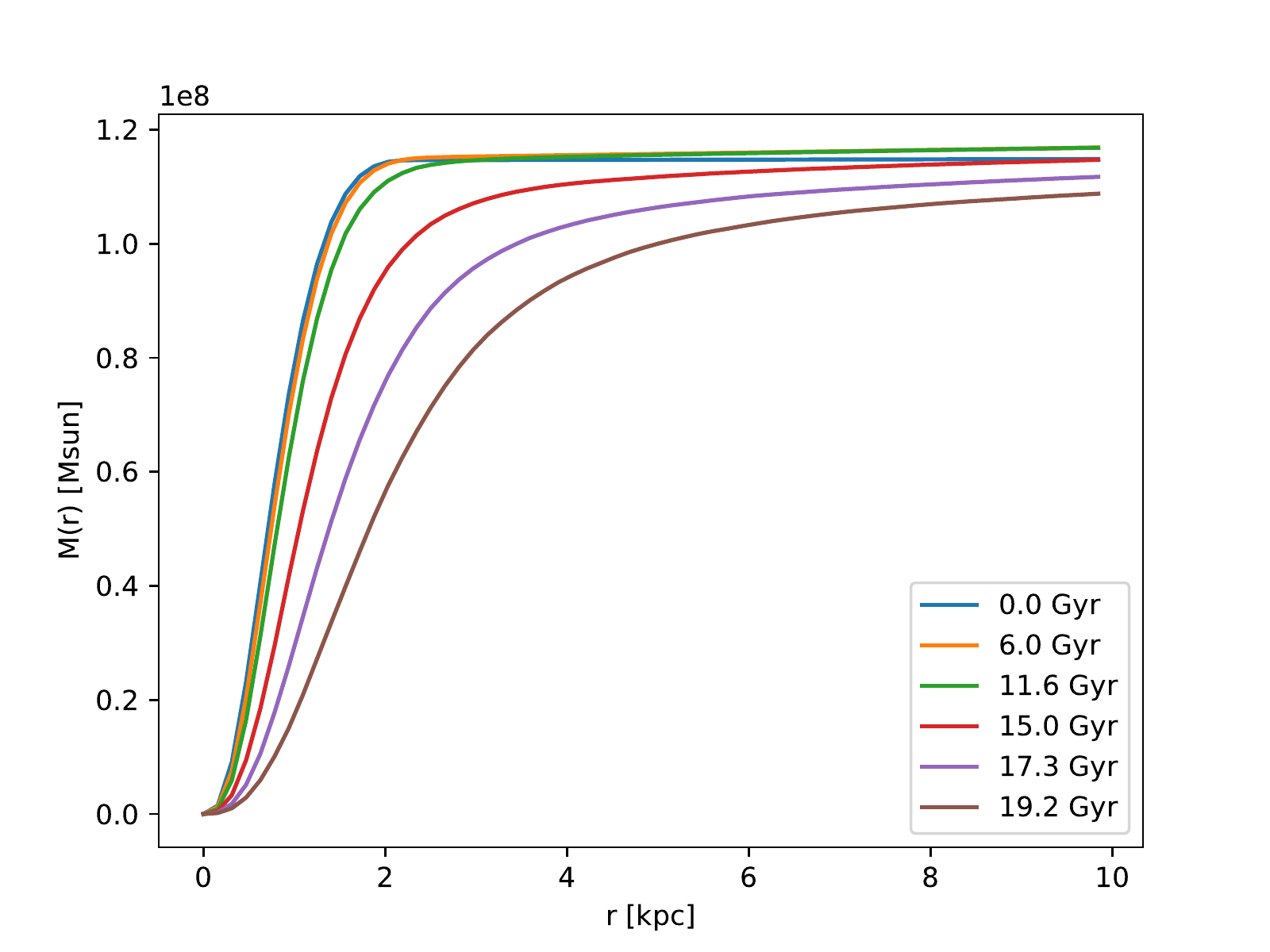} &
        \includegraphics[width=0.47\textwidth]{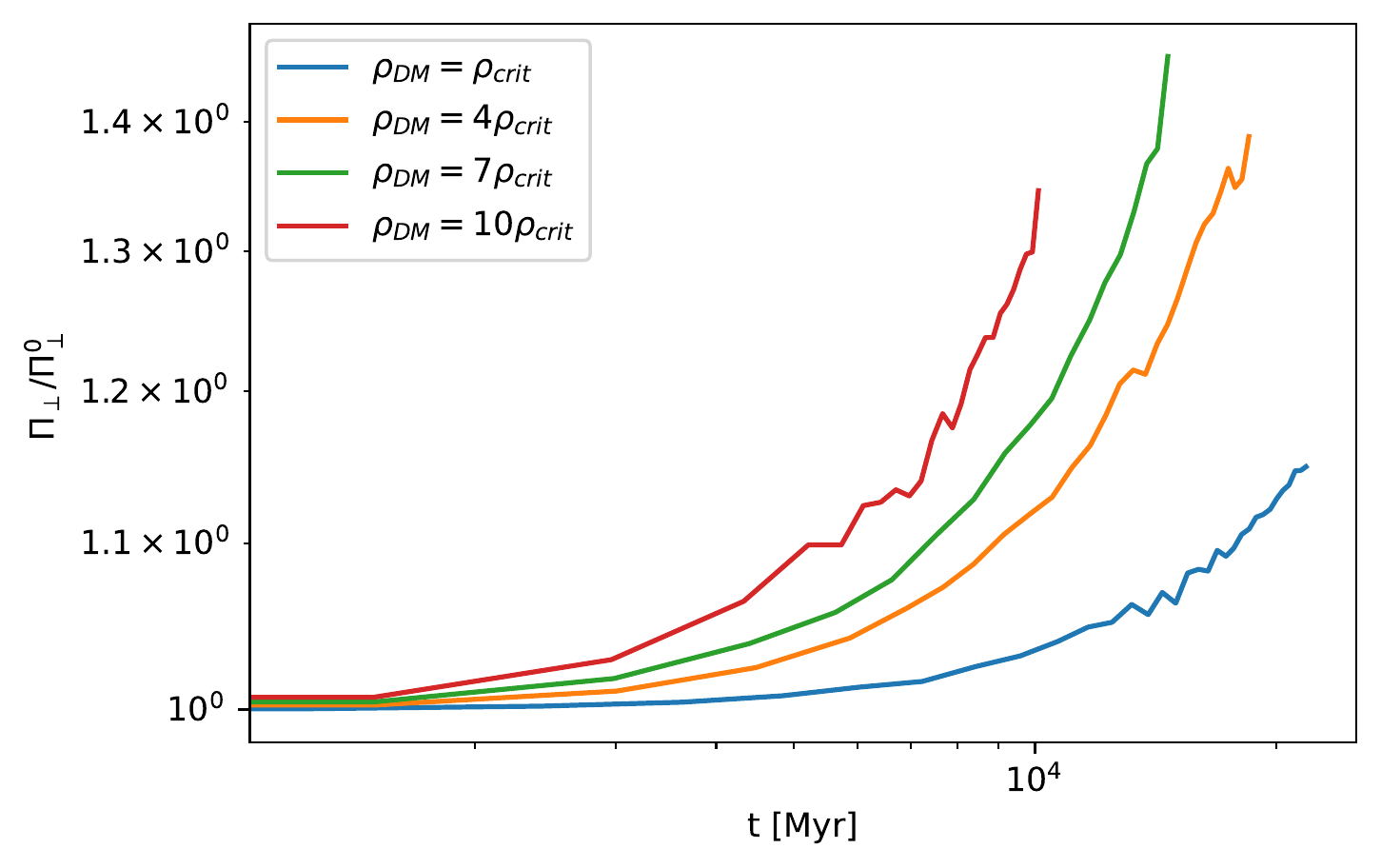}
        \end{tabular}
        \caption{\label{fig:insta} Left panel : Time evolution of the the mass enclosed in spheres of radius $r$. While the initial condition represents the standard King profile, the deformation of this profile characterizes the dynamical instability proper to the dipolar DM model. Right panel : the time evolution of $\Pi(r=1.7 \text{ kpc})$ for different DM densities. We checked that the scaling of the typical time is as expected $\propto 1/\sqrt{\rho_{\text{dm}}}$.}
\end{figure*}

Another way to apprehend this instability is to consider the mass enclosed in a sphere centered at the center of the galaxy. We show it in the left panel of figure \ref{fig:insta}. The initial line corresponds to a King profile distribution and as time passes, the distribution evolves to a very dissolved galaxy.

\subsection{Characteristic time of the instability}
We now elaborate a bit more on the dipolar DM instability. Physically, the dark matter is sourced by $\vec{g}$ and clumps into structures, the dipolar dark matter is at rest: $\vec{v}=\vec{0}$ because the internal force $\vec{F}_\text{int}$ exactly compensates the gravitational force $\vec{g}$. The polarization $\vec{\Pi}$ however only feels the internal force $\vec{F}_\text{int}$ that will diverge within the free-fall time of the dark matter. Working at leading order and replacing $\vec{F}_\text{int}$ by its expression from eqs.~\eqref{eq:Fint}, \eqref{eq:Pitog} and \eqref{eq:Taylor_g} in terms of $\vec{\Pi}$, Eq.~\eqref{eq:dynPi} then reads:
\be
\label{eq:instaPi}
\frac{\text{d}^2 \vec{\Pi}}{\text{d}t^2} \simeq 4 \pi G \rho_{\rm dm} \vec{\Pi},
\ee
leading to exponential solutions with a characteristic time $\tau \simeq \sqrt{\pi/G\rho_{\rm dm}}$m \citep[][]{Blanchet_2009}, of the order of the free-fall time. 

In the suite of simulations $\text{K}_{1}$, $\text{K}_{4}$, $\text{K}_{7}$, $\text{K}_{10}$, we varied the density of the dipolar DM halo to measure the scaling of the instability, and we show the results in the right panel of figure  \ref{fig:insta}. As expected, by fitting our results to an exponential, we did find the scaling $1/\sqrt{\rho_{\rm dm}}$. We check the numerical convergence of this result by performing a low resolution simulation of $K_1$ by multiplying the mass of the baryons particles by a factor 2 (and, hence, dividing the number of baryons by a factor 2). The time evolution of $\Pi$ agrees to less than $1 \%$ thus validating our choice of resolution. 

It is also interesting to note that the presence of the instability of the dipolar DM confirms a posteriori the necessity of the weak clustering hypothesis, and means that the density of dipolar DM particles is necessarily close to the critical density. This is of relevance for direct detection of such dipolar DM particles, which would have a very low density in the Solar neighbourhood.

\section{Conclusions and perspectives}
\label{sec:ccl}
We performed for the first time N-body experiments within the dipolar DM scenario proposed in \citet{Blanchet_2009}. In this article, paving the way for more involved setups, we focused on the galactic realization of dipolar DM. We reviewed the relevant analytical results in section \ref{sec:Theo} and showed how we implemented them in \texttt{RAMSES} in section \ref{sec:num}. We restricted ourselves to spherical symmetric configurations as well as the low acceleration (`deep-MOND') regime. We explicitly gave in section \ref{sec:pot} the exact form of the potential in this set-up. 

We studied the equilibrium solution of a low-density static spherically symmetric DM fluid with $\vec{v}=\vec{0}$. This equilibrium configuration relies on the weak clustering hypothesis, namely that once the medium polarizes itself in the non-linear regime of structure formation, the internal force counteracts gravity, and allows most particles not at rest to escape, thereby leaving only a very low monopolar density almost at rest with respect to its own frame.

After carefully checking that when freezing the baryons and freezing the DM fluid and its polarization separately, our galaxy was stable and recovering the expected gravitational field, we ran live simulations with different densities for the DM halo, and found back the instability analytically expected from \citet{Blanchet_2009}, with a characteristic time $\tau \propto 1/\sqrt{\rho_{\rm dm}}$. This dynamical instability at the level of the polarization vector translates into a dissolution of the galaxies that we simulated over cosmological time-scales. We explored how they dissolved with the cumulated mass profile and the scaling of this instability with respect to the density of DM, confirming the analytical expectations. This thus validates our code whilst independently confirming from a numerical perspective previous analytical computations. 

We explicitly show how our simulated galaxies naturally remain on the Radial Acceleration Relation over time within the Dipolar DM framework. As the dynamical instability develops, this results breaks down and we show how the RAR gets destroyed, starting from the center of our galaxies where the Newtonian gravitational field was the strongest.

Our work opens new perspectives to test the dipolar DM framework. A natural continuation will be to investigate setups where two galaxies such as the one considered in the article are merging. The main perspective for this model will then be to go cosmological and perform a zoom simulation where both cosmological and galactic dynamics coexist. There, one fascinating feature of the dipolar DM is the presence of a non-gaussianities growing with time \cite{Blanchet:2012ey}. This very rare type of non-gaussianities could be a smoking-gun of this model but also a way to constrain the form of the potential of such a model.

\section*{Acknowledgements}
The authors thank the referee for a thoughtful and constructive report. CS, BF and RI acknowledge funding from the European Research Council (ERC) under the European Union's Horizon 2020 research and innovation program (grant agreement No.\ 834148).
This work has made use of the Infinity Cluster hosted by the Institut d'Astrophysique de Paris.
The analysis was partially made using the YT python package \cite{Turk:2010ah}, as well as IPython \cite{Perez:2007emg}, Matplotlib \cite{Hunter:2007ouj}, NumPy \cite{vanderWalt:2011bqk} and SciPy \cite{Virtanen:2019joe}. 

\section*{Authors' Contribution}
Based on an original idea put forward by BF and RI, CS implemented in \texttt{RAMSES} the dipolar dark matter equation helped by YD. The simulations presented in this work were performed by CS using initial data provided by GT. CS and BF drafted the manuscript and all authors improved it heavily by their comments.

\section*{Data Availability}
The data supporting the findings of this research is available upon reasonable demand to the corresponding author. The patch of the \texttt{RAMSES} code will be soon publicly available through the GitHub
repository: \url{https://github.com/cspotz/}.

\section*{Carbon Footprint}
Following \cite{berthoud:hal-02549565} to convert\footnote{Including the global utilisation of the cluster and the pollution due to the electrical source, the conversion factor is 4.7 gCO2e/h core} the number of CPU hours required to obtain the data for this work, we have used 0.47 tCO2eq.


\bibliographystyle{mnras}
\bibliography{biblio} 





\bsp        
\label{lastpage}
\end{document}